\documentclass[12pt,aps,floats]{revtex4}
\usepackage{amsmath}
\usepackage{amssymb}
\usepackage{mathtools}

\begin{document}
\def\bigint{{\displaystyle\int}}
\def\simlt{\stackrel{<}{{}_\sim}}
\def\simgt{\stackrel{>}{{}_\sim}}

\title{Comment on 'White, T., Mutus, J., Dressel, J. et al., "Preserving entanglement during weak measurement demonstrated with a violation of the Bell-Leggett-Garg inequality", npj Quantum Information 2, 15022 (2016).'}
%\title{Unraveling quantum entanglement without disentangling it, with weak measurements}
%\title{Unraveling the entanglement without disentangling it, with the help of weak measurements}

\author{David H. Oaknin}
\affiliation{Rafael Ltd, IL-31021 Haifa, Israel, \\
e-mail: { d1306av@gmail.com}}

\begin{abstract}
In reference \cite{White} experimental results were presented that clearly prove that the quantum entanglement between two qubits is preserved after weak enough measurements are performed on them. The theoretical interpretation of the reported results, however, requires further consideration. The remarks made in this paper may have serious implications both for quantum foundations and for quantum cryptography.
\end{abstract}

\maketitle

\noindent
In the experiment reported in \cite{White,Dressel} sequences of pairs of entangled superconducting qubits prepared in an entangled Bell's state were weakly coupled to pairs of ancilla qubits, before the tetrads were projected through strong measurements along predetermined orientations. This innovative experimental setting purportedly allows to measure for each tetrad the four outcomes that are required to build the CHSH correlator and, thus, it supposedly allows to bypass the so-called 'disjoint sampling loophole' \cite{Larsson} and the 'clumsiness loophole' \cite{Wilde} of the standard setting of Bell's experiment, in which only two of the four required outcomes are measured in every single realization. 

The collected experimental data reported in \cite{White}, see Fig.1, clearly demonstrates that the correlator violates the CHSH inequality and, thus, it proves that the entanglement between the two original qubits  - or Bell qubits, as they are referred to in  \cite{White,Dressel} - is preserved in spite of their weak interaction with the ancilla qubits. Thus, it supposedly buries all hopes to build a successful local model of hidden variables for the entangled Bell's states exploiting the loopholes associated to the fact that only two of the four outcomes involved in the CHSH correlator are measured in every single realization of a Bell experiment \cite{Hess1}. However, as we shall now show, this last conclusion cannot be properly justified.      

In the experiment reported in \cite{White} only the outcomes obtained by strongly measuring the polarizations of the two Bell qubits are indeed binary, $\beta_{1,2}=\pm 1$, while the outcomes obtained by projecting the two ancilla qubits are not, $\alpha_{1,2}=\pm 1/V$: they get normalized by the strength $V \rightarrow 0^+$ of their coupling to the two original qubits and, moreover, are supposedly contaminated by noise produced by the detectors \cite{Dressel}. Therefore, the analysis presented in \cite{White} of their collected experimental data relies on the assumption that this noise is unbiased \cite{Dressel}, so that over a long enough sequence of repetitions of the experiment the effect of noise gets averaged out and the measured correlations, after proper normalization, are indeed associated to underlying binary signals. This assumption is the main interest of this note, because as we shall now show it is straightforward to prove that it cannot be correct.

The assumption seems to be trivially justified, since the detectors were previously calibrated with single qubits and the noise that affects the outcomes of the measurements performed on each one of the ancilla qubits should not be correlated if the measurements are causally disconnected. However, as we have noticed above the collected data shows a clear violation of the CHSH inequality and it is straightforward to prove, see the straightforward theorem below, that there cannot exist a sequence of four binary outcomes $\left\{V\cdot\alpha_{1,2}(n), \ \beta_{1,2}(n)\right\}_{n \in N}$ whose correlations violate the CHSH inequality. That is, the violation of the CHSH inequality reported in \cite{White} cannot be accounted for by any model whose prediction for the possible outcomes of the experiment consists of  binary 4-tuples contaminated only by unbiased noise \cite{Dressel}. In other words, the violation of the CHSH inequality reported in \cite{White} is necessarily associated to the apparent {\it noise} in the collected data for the outcomes $V \cdot \alpha_{1,2}$, which strongly suggest that some component of that {\it noise} is not actually noise. In fact, the experimental results reported in \cite{White} can be reproduced within the framework of a model of local hidden variables \cite{oaknin1,oaknin2,oaknin3}.
\\
% A possible expplanation to these results is provided by the model of hidden variables discussed in \cite{oaknin1,oaknin2}.

\begin{figure}
\begin{center}
\includegraphics[width=21cm]{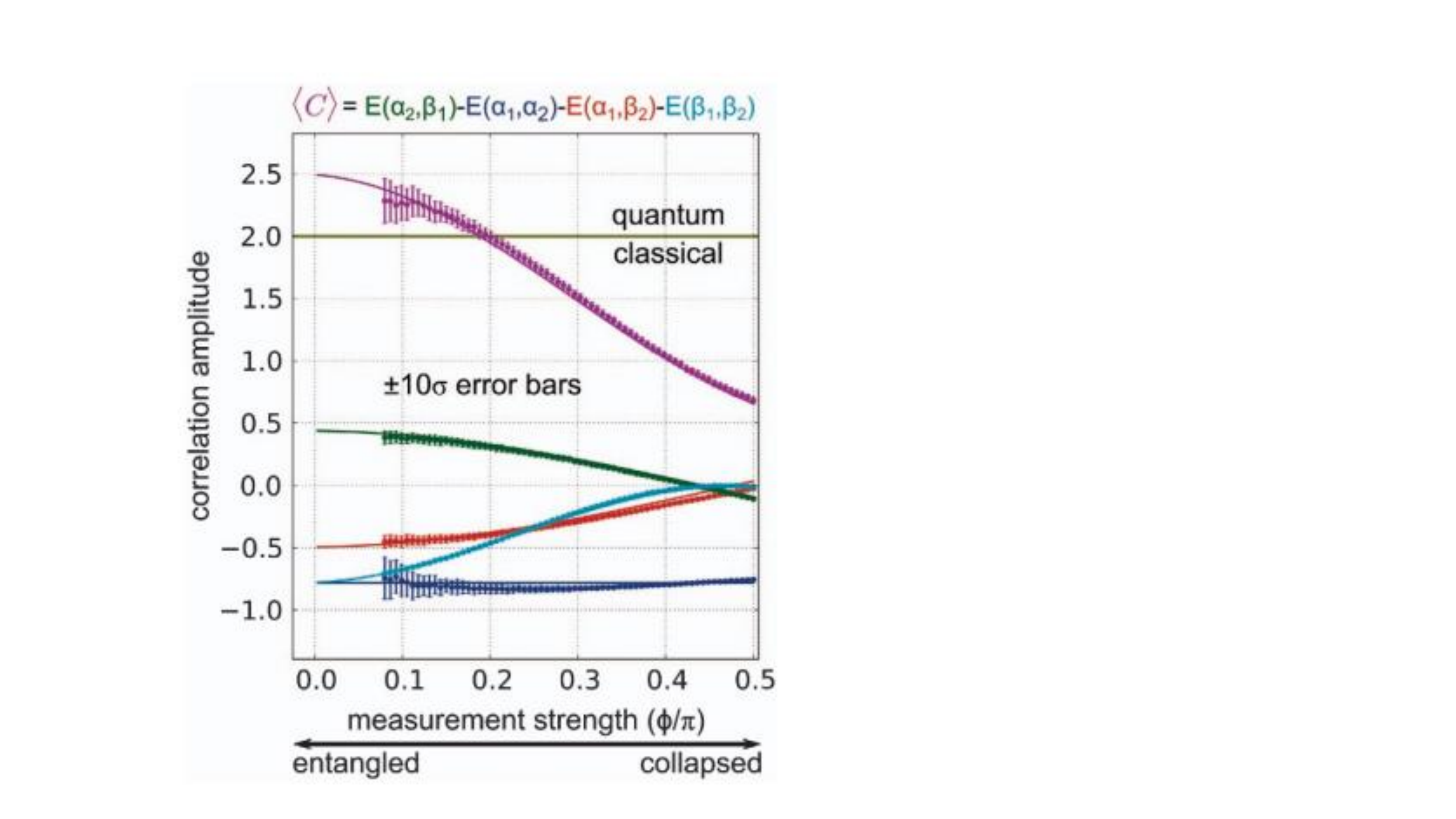}
\end{center}
\caption{Figure reproduced from White {\it et al.} \cite{White}, figure 2. The figure shows the measured correlations between polarization measurements performed on the Bell and ancilla qubits along relative directions that correspond to those used in usual Bell experiments to obtain a maximal violation of the CHSH inequality. Indexes $\beta_1, \ \beta_2$ denote the Bell qubits and $\alpha_1, \ \alpha_2$ the ancilla qubits, while E denotes the correlation between the outcomes of the polarization measurements performed on them. Retrieved from  https://www.nature.com/articles/npjqi201522. }
\label{fig:fig1}
\end{figure}

Finally, the experimental results reported in \cite{White} clearly prove that the Bell qubits remain entangled after coupling them very weakly to the ancilla qubits, and yet the outcomes of the projective measurements performed on the ancilla qubits are quite strongly correlated to the outcomes of the projective measurements performed on the Bell qubits. These results suggest that it might be possible to predict with high probability the binary outcomes of the projective measurements that shall be performed on the two entangled Bell qubits by previously weakly coupling them to two ancilla qubits and performing on the later a long sequence of very weak measurements instead of single projective measurements \cite{Oreshkov}. 

Indeed, the preservation of the entanglement of the Bell qubits after weakly interacting with the ancilla qubits could be confirmed by performing on the former a standard Bell experiment. In particular, it would be of great interest to perform a standard Bell experiment on the Bell qubits after strongly projecting the ancilla qubits along exactly the same directions that will be later on tested on the Bell qubits. Moreover, the projection of the ancilla qubits could be done through a long sequence of very weak measurements according to the protocol discussed in \cite{Piacentini}, so that an average value can be obtained for every single ancilla qubit. The average values obtained for each pair of ancilla qubits are expected to be strongly correated to the outcomes of the projective measurements performed on the pair of entangled Bell qubits later on. In other words, the average values obtained for the ancilla qubits may allow to predict with high confidence the outcomes of the projective measurements not yet performed on the pair of entangled Bell qubits. Such a protocol could have very serious practical implications regarding quantum communication protocols \cite{Scarani} and also in the study of the theoretical foundations of quantum mechanics \cite{Weinberg}.

{\bf Theorem:} For any sequence of 4-tuples $\left\{\left(a_1(n), a_2(n), b_1(n), b_2(n)\right)\right\}_{n \in \{1,2,...,N\}}$ of binary numbers $a_1(n), a_2(n), b_1(n), b_2(n) \in \left\{-1,+1\right\}, \ \ \forall n \in \{1,2,...,N\}$, the following inequality holds:
\begin{equation}
\frac{1}{N} \left|\sum_{n} a_1(n) \cdot b_1(n) + \sum_{n} a_1(n) \cdot b_2(n) + \sum_{n} a_2(n) \cdot b_1(n) - \sum_{n} a_2(n) \cdot b_2(n)\right| \le 2.
\end{equation}   

Proof:
\begin{eqnarray*}
a_1(n) \cdot b_1(n) + a_1(n) \cdot b_2(n) + a_2(n) \cdot b_1(n) - a_2(n) \cdot b_2(n) = \\
= a_1(n) \cdot \left(b_1(n) + b_2(n)\right) + a_2(n) \cdot \left(b_1(n) - b_2(n)\right) = \pm 2.
\end{eqnarray*} 
\\

The author declares that he has not any competing financial or non-financial interests beyond his purely scientific interests.

\end{document}